\documentclass[12pt]{article}
\usepackage[margin=1.0in]{geometry}
\usepackage{setspace}
\usepackage{blindtext}
\usepackage{amsmath}
\usepackage{amsthm,amssymb}
\usepackage{amstext,verbatim}
\usepackage{amsthm}
\usepackage{bm}
\usepackage{amssymb,amsmath,mathrsfs,dsfont,setspace}
\usepackage{graphicx}
\usepackage{array}
\usepackage{natbib}
\usepackage{epsfig}
\usepackage{amssymb}
\usepackage{sectsty}
\usepackage{algorithm}
\usepackage{graphicx,psfrag,subfigure,amsmath,amssymb,appendix,multirow,float,threeparttable,abstract}
\usepackage{mathrsfs,setspace}
\usepackage{color}
\usepackage{longtable}
\usepackage{pifont}
\usepackage{rotating}
\usepackage{adjustbox}
\usepackage{appendix}
\usepackage{verbatim}
\usepackage{datetime}

\fontsize{12}{0}
\sectionfont{\large}
\subsectionfont{\large}
\subsubsectionfont{\large}
\numberwithin{equation}{section}
{\begin{center} \begin{tabular}{|@{\hspace{.15in}}c@{\hspace{.15in}}|}
			\hline \\ \begin{minipage}[t]{\boxedparwidth}
				\setlength{\parindent}{.25in}}%
			{\end{minipage} \\ \\ \hline \end{tabular} \end{center}}

\newtheorem{theorem}{Theorem}
\newtheorem{assumption}{Assumption}
\newtheorem{lemma}{Lemma}
\newtheorem{proposition}{Proposition}
\newtheorem{corollary}{Corollary}
\newtheorem{definition}{Definition}
\newtheorem{remark}{Remark}

\newcommand{\ehg}{\widehat{\mathcal E}_{\gamma_n}}
\newcommand{\ehag}{\widehat{\mathcal E}_{j,\gamma_n}}

\newcommand{\Sign}{\mathrm{sign}}

\newcommand{\Rank}{\mathrm{rank}}

\newcommand{\vartauj}{\widehat{\omega}_{jj'}^2}

\newcommand{\sdtauj}{\widehat{\omega}_{jj'}}

\title{Sure Screening for Transelliptical Graphical Models}
\author{Yuxiang Xie, Chengchun Shi, Rui Song}
\date{January 15, 2016}

\newcommand{\Mean}{{\mathbb{E}}}

\begin{document}
\maketitle
\doublespacing

\begin{abstract}
\indent \indent We propose a sure screening approach for recovering the structure of a transelliptical graphical model in the high dimensional setting. We estimate the partial correlation graph by thresholding the elements of an estimator of the sample correlation matrix obtained using Kendall's tau statistic. Under a simple assumption on the relationship between the correlation and partial correlation graphs, we show that with high probability, the estimated edge set contains the true edge set, and the size of the estimated edge set is controlled. We develop a threshold value that allows for control of the expected false positive rate. In simulation and  on an equities data set, we show that transelliptical graphical sure screening performs quite competitively with more computationally demanding techniques for graph estimation. \\
\noindent {\small \textit{some key words:} High Dimensionality;  Kendall's tau; Partial Correlation;  Sparsity; Undirected Graph}

\end{abstract}

\section{Introduction} \label{intro}

Consider the random vector $X=\left(X_1,\ldots,X_p\right)^T$, and an undirected graph denoted by $G=\left(\mathcal{V}, \mathcal{E}\right)$, where $\mathcal{V} = \left\{1,\dots,p\right\}$ is the set of nodes, and $\mathcal{E}$ is the set of edges describing the conditional dependence relationships among $X=\left(X_1,\ldots,X_p\right)^T$. A pair $\left(j,j'\right)$ is contained in the edge set $\mathcal{E}$ if and only if $X_j$ is conditionally dependent on $X_{j'}$, given all remaining variables $X_{\mathcal{V}\backslash\{j,j'\}} = \left\{ X_k; k \in \mathcal{V}\backslash\left\{j,j'\right\} \right\}$.

In recent years, many methods have been developed to recover the structure of  graphical models in the high dimensional setting. Many authors have studied the Gaussian graphical model, in which conditional dependence is encoded by the sparsity pattern of the inverse covariance matrix \citep[and references therein]{yl07, fr08, ro08, ra09}. \cite{liu09,liu12a}  introduced the  nonparanormal distribution, which results from univariate monotonic transformations of the Gaussian distribution, and showed that the structural properties of the inverse covariance matrix of the Gaussian distribution carry over to the corresponding nonparanormal distribution. \cite{ra10} and \cite{an12} considered recovering the structure of an Ising graphical model. \cite{NIPS2012_0659}  studied a class of graphical models in which the node-wise conditional distributions arise from the exponential family. Moreover, \cite{yang14} and \cite{chen15} considered the problem of structure recovery for mixed graphical models.

\cite{luo12} proposed a computationally-efficient screening approach for Gaussian graphical models, which they called \emph{graphical sure screening} (GRASS). They estimated an edge between the $j$th and  $j'$th nodes if the sample correlation between the $j$th  and $j'$th features exceeds some threshold $\gamma$. Then the $j$th node's estimated neighborhood    contains the true neighborhood with very high probability, under certain simple assumptions. GRASS requires only $\mathcal{O}(p^2)$ operations, while most other existing methods for estimating the graph require $\mathcal{O}(p^3)$ computations \citep{fr08}. 

However, GRASS requires that the data follows a multivariate normal distribution. In many settings, a reliance on exact normality is not desirable  \citep{liu09,liu12a, liu12b}. In this paper, we propose \emph{transelliptical GRASS}, an extension of the GRASS procedure to the \textit{transelliptical graphical model} family, introduced by \cite{liu12b}. We show that under a certain set of simple assumptions, the desirable statistical properties held by GRASS are also held by transelliptical GRASS. However, due to the relaxation of multivariate normality, the graph that we estimate represents the partial correlation, rather than conditional dependence, among variables. 
 
The rest of this paper is organized as follows. In Section~\ref{trans_kendall}, we provide some background on transelliptical graphical models, and present a useful property of Kendall's tau statistic. In Section~\ref{tgrass}, we establish the theoretical properties for transelliptical sure screening, which include the sure screening property, size control of the selected edge set, and the control of the expected false positive rate. Furthermore, we provide a choice of the threshold value that leads to the aforementioned desirable properties. Simulation studies are presented in Section~\ref{sec:sim}, and an application  to an equities data set is shown in Section~\ref{equities}. We close with a discussion in Section~\ref{discussion}.

\section{Preliminaries} \label{trans_kendall}
The transelliptical distribution \citep{liu12b} is a generalization of the nonparanormal distribution \citep{liu09, liu12a}. The transelliptical distribution extends the elliptical distribution in much the same way that the nonparanormal extends the normal distribution. We first provide the definition of the elliptical distribution. 
 \begin{definition} \label{ellip_dis}
	Let $\mu \in \mathbb{R}^p$ and $\Sigma \in \mathbb{R}^{p\times p}$ with $\Rank\left(\Sigma\right)=q\le p$. A $p$-dimensional random vector $X$ has an elliptical distribution, denoted by $X\sim \mbox{EC}_p\left(\mu, \Sigma, r\right)$, if it has a stochastic representation 
	\begin{eqnarray}\label{ellip}
	X\stackrel{d}{=}\mu+r AU,
	\end{eqnarray}
	where $U$ is a random vector uniformly distributed on the unit sphere in $\mathbb{R}^q$, $r \ge 0$ is a scalar random variable independent of $U$, and $A \in \mathbb{R}^{p\times q}$ satisfies $\Sigma = AA^{T}$.
	\end{definition}
	 In Definition~\ref{ellip_dis},  $X\stackrel{d}{=}Y$ indicates that $X$ and $Y$ have the same distribution. 
%
%
	Many multivariate distributions, such as the multivariate normal and the multivariate t-distribution,  belong to the elliptical distribution family. 
	

From now on, we assume that $A$ in (\ref{ellip}) is a full rank $p\times p$ matrix so that $\Sigma$ is positive definite. In addition, assume that $\Sigma$ has unit diagonal elements. We let $\rho_{jj'}$ denote the $\left(j,j'\right)$th entry of $\Sigma$. We also assume that the scalar random variable $r$ in (\ref{ellip}) has density $g(\cdot)$ and $\Mean(r^2)<\infty$. 

\begin{definition} \label{transellip_dis}
	A continuous random vector $X=\left(X_1,\ldots,X_p\right)^T$ follows a $p$-dimensional transelliptical distribution, denoted by $\mbox{TE}_p\left(\Sigma, r;f_1,\dots,f_p\right)$, if there exist monotone univariate functions $\left\{f_j\right\}_{j=1}^p$ and a nonnegative random variable $r$ satisfying $\Pr(r=0)=0$, such that  $Z \equiv f(X) \equiv \left(f_1\left(X_1\right),\ldots,f_p\left(X_p\right)\right)^T \sim \mbox{EC}_p\left(0, \Sigma, r\right)$. We refer to $Z=\left(Z_1,\ldots,Z_p\right)^T=\left(f_1\left(X_1\right),\ldots,f_p\left(X_p\right)\right)^T$ as the latent variables of $X=\left(X_1,\ldots,X_p\right)^T$. 
\end{definition}

\begin{remark} \label{rm22}
	A random vector $X=\left(X_1,\ldots,X_p\right)^T$ follows a nonparanormal distribution \citep{liu09, liu12a} if there exist monotone univariate functions $\left\{f_j\right\}_{j=1}^p$ such that $Z=f(X)=\left(f_1\left(X_1\right),\ldots,f_p\left(X_p\right)\right)^T \sim N_p({0},{\Sigma})$. Therefore, the transelliptical distribution is a strict extension of the nonparanormal distribution.
\end{remark}

Given a transelliptical distribution $\mbox{TE}_p\left(\Sigma, r;f_1,\dots,f_p\right)$, we can define an undirected graph $G=\left(\mathcal{V}, \mathcal{E}\right)$, where $\mathcal{V}=\left\{1,\ldots,p \right\}$, and $\left(j,j'\right) \in \mathcal{E}$  if and only if $\left(\Sigma^{-1}\right)_{jj'} \neq 0$ --- that is, if and only if the latent variables $Z_j=f_j(X_j)$ and $Z_{j'}=f_{j'}(X_{j'})$ are partially correlated \citep{liu12b}. In the special case of a nonparanormal distribution, a zero entry of the precision matrix $\Sigma^{-1}$ further implies conditional independence between the corresponding pair of  random variables.

Next, we present the definition and some theoretical properties of the Kendall's tau statistic.
\begin{definition} \label{kend_tau}
	Given $n$ independent draws from $X=(X_1,\ldots,X_p)^T$, such that $X_{ij}$ is the value of the $j$th variable in the $i$th observation, the population-level Kendall's tau statistic between $X_j$ and $X_{j'}$ is defined as
	\begin{equation}
		\tau_{jj'}=\Mean \left[\Sign \left( \left(X_{1j}-X_{2j}\right)\left(X_{1j'}-X_{2j'}\right)\right)\right].
	\end{equation}
	The sample estimator of Kendall's tau is defined as 
	\begin{equation} \label{kendalltau}
	\widehat{\tau}_{jj'}=\frac{2}{n(n-1)}\sum_{1\leq i\leq i'\leq n}\Sign \left( \left(X_{ij}-X_{i'j}\right)\left(X_{ij'}-X_{i'j'}\right)\right).
	\end{equation}
\end{definition}
 Now suppose that we have $n$ independent draws from $X=(X_1,\ldots,X_p)^T \sim \mbox{TE}_{p}(\Sigma, r;f_1,\dots,f_p)$, such that $X_{ij}$ is the value of the $j$th variable in the $i$th observation. Then there is a simple connection between the population-level Kendall's tau statistic $\tau_{jj'}$ and $\rho_{jj'}$. 
\begin{lemma} \label{lemmaliu}
	\citep{liu12b} For  $X=\left(X_1,\ldots,X_p\right)^T \sim \mbox{TE}_{p}(\Sigma, r;f_1,\dots,f_p)$, we have $\rho_{jj'}=\sin(\frac{\pi}{2}\tau_{jj'})$.
\end{lemma}

Lemma~\ref{lemmaliu} motivated
\cite{liu12b} to estimate $\rho_{jj'}$ using

\begin{equation} \label{stau} 
\widehat{S}^{\tau}_{jj'} = \begin{cases}
\sin\left(\frac{\pi}{2}\widehat{\tau}_{jj'}\right) & \text{if $j\neq j'$}\\ 
1 & \text{if $j=j'$} 
\end{cases}. 
\end{equation}



\section{Transelliptical Graphical Sure Screening} \label{tgrass}
\subsection{Proposed Approach} \label{tgrass:approach}
Suppose that we have $n$ independent draws from $X=(X_1,\ldots,X_p)^T \sim \mbox{TE}_{p}(\Sigma, r;f_1,\dots,f_p)$. We define  $\mathcal{E}\equiv \left\{\left(j,j'\right): j < j', \left(\Sigma^{-1}\right)_{jj'} \neq 0\right\}$ to be the true edge set, and $\mathcal{E}_{j}\equiv \left\{ j': j'\neq j, \left(\Sigma^{-1}\right)_{jj'}\neq 0 \right\}$ to be the true neighborhood for the $j$th node. We propose to estimate $\mathcal{E}$ and $\mathcal{E}_{j}$ as follows,
\begin{equation}\label{edgeset}
\ehg=\left\{\left(j,j'\right):j<j',\left|\widehat{S}^{\tau}_{jj'}\right|>\gamma_{n,jj'}\right\} 
\end{equation}
and
\begin{equation}\label{neigh}
\ehag=\left\{j':j'\neq j,\left|\widehat{S}^{\tau}_{jj'}\right|>\gamma_{n,jj'}\right\},	
\end{equation}
where $\gamma_{n,jj'}>0$ is some threshold value that we will specify in the following sections, and $\widehat{S}^{\tau}_{jj'}$ is defined in \eqref{stau}. We refer to $\widehat{\mathcal{E}}_{\gamma_n}$ and $\widehat{\mathcal{E}}_{j,\gamma_n}$ as the \emph{transelliptical graphical sure screening} (transelliptical GRASS) estimators.

\subsection{Theoretical Properties} \label{tgrass:theory}
\indent \indent We now present some theoretical properties of transelliptical GRASS. Proofs are in the Appendix.   

\begin{assumption} \label{min_rho}
	For some constant $C_1>0$ and $0<\kappa<1/2$,
	\begin{equation*}
	\min_{\left(j,j'\right)\in\mathcal{E}}\left|\rho_{jj'}\right|\ge C_1 n^{-\kappa}.
	\end{equation*}
\end{assumption}
 Assumption 1 requires that the elements in the edge set $\mathcal{E}$ correspond to sufficiently large values in the correlation matrix.  We next present the sure screening property in Theorem 1.

\begin{theorem} \label{control_fnr}
	Suppose that Assumption~\ref{min_rho} holds, and that $\log(p)=C_3 n^{\xi}$ for some constants $C_3>0$ and $\xi\in(0,1-2\kappa)$. Let $\gamma_{n,jj'} \equiv \gamma_{n}=\frac{2}{3}C_1 n^{-\kappa}$. 
	Then there exist constants $C_4$ and $C_5$ such that 
	\begin{align*}
	\Pr(\mathcal{E}\subseteq\ehg) & \geq 1-C_4\exp(-C_5n^{1-2\kappa})
	\end{align*}
	and
	\begin{align*}
	\Pr(\mathcal{E}_j\subseteq\ehag) & \geq 1-C_4\exp(-C_5n^{1-2\kappa}).
	\end{align*}
	Conversely, if $\min_{(j,j')\in \mathcal{E}}|\rho_{jj'}|< C_1n^{-\kappa}/3$, then there exist constants $C_6$ and $C_7$ such that
	\begin{eqnarray}\label{thm1conclusion2}
	\Pr(\mathcal{E}\not \subseteq \ehg) {\geq} 1 - C_6\exp(-C_7n^{1-2\kappa}).
	\end{eqnarray}
\end{theorem}

Theorem~\ref{control_fnr} guarantees that the candidate edge set obtained from transelliptical GRASS contains the true edge set with high probability, which means the screening method will not result in false negatives with high probability. Moreover, (\ref{thm1conclusion2}) suggests that Assumption 1 is necessary up to a constant. The following corollary shows that under Assumption~\ref{min_rho}, transelliptical GRASS can recover the connected components of $\mathcal{E}$ with high probability.

\begin{corollary}\label{concomp}
	Suppose there are $h$ connected components in the graph $\mathcal{E}$, and that the $l$th connected component contains the variables
	$x_1^{(l)}, \dots, x_{p_l}^{(l)}$, where $ \sum_{l=1}^h p_l=p$.
	That is, $x_{j}^{(s)}$ and $x_{j'}^{(t)}$
	are partially uncorrelated for $s \not= t$. Suppose Assumption \ref{min_rho} and the conditions in Theorem~\ref{control_fnr} hold. Let $\gamma_{n,jj'} = \gamma_n = 2C_1 n^{-\kappa}/3$. Then
	the connected components of $\ehg$ are the same as the connected components of $\mathcal{E}$ with probability at least $1 - C_4\exp\left( -C_5n^{1-2\kappa} \right)$.
\end{corollary}

Our next theorem will provide a bound on the size of $\ehag$. This requires an additional assumption.

\begin{assumption} \label{max_eigen}
	There exist constants $\alpha\geq0$ and $C_{2}>0$ such that $\Lambda_{\max}(\Sigma)\leq C_{2}n^{\alpha}$, where $\Lambda_{\max}(\Sigma)$ is the largest eigenvalue of $\Sigma$.
\end{assumption}
Assumption~\ref{max_eigen} allows the largest eigenvalue of the population covariance matrix $\Sigma$ to diverge as $n$ grows.

\begin{theorem} \label{control_size}
	Let $\gamma_{n,jj'} = \gamma_{n}=\frac{2}{3}C_1 n^{-\kappa}$. 
	 Under Assumptions~\ref{min_rho}--\ref{max_eigen}, if $\log\left(p\right)=C_3 n^{\xi}$ for some constants $C_3>0$ and $\xi\in \left(0,1-2\kappa\right)$, then $\Pr \left(  \left|  \ehag \right| \leq O \left( n^{2\kappa+\alpha} \right) \right)$$\geq 1-C_4\exp\left(-C_5n^{1-2\kappa}\right)$, where the constants $C_4$ and $C_5$ are as in Theorem~\ref{min_rho}.
\end{theorem}

Next we propose a choice of the threshold $\gamma_{n,jj'}$ that enables us to control the expected false positive rate, defined as ${|\ehg\cap\mathcal{E}^c|}/{\mathcal{E}^c}$, at a pre-specified value. Here, $\mathcal{E}^c \equiv \left\{\left(j,j'\right): j < j', \left(\Sigma^{-1}\right)_{jj'} = 0\right\}$, and $\ehg$ is defined in \eqref{edgeset}. This requires an additional assumption.

\begin{assumption} \label{max_rho}
	For the same $\xi$ as in Theorem~\ref{control_fnr}, 
	\begin{equation*}
	\max_{(j,j')\notin\mathcal{E}}\left|\rho_{jj'}\right|=o\left(n^{-\frac{1-\xi}{2}}\right).
	\end{equation*}
\end{assumption}

\begin{theorem} \label{control_fpr}
	Suppose that Assumptions 1--3 hold. If $\log(p)=C_3n^{\xi}$ for $\xi$ defined in Theorem~\ref{control_fnr}, then we can control the asymptotic expected false positive rate at $f/|\mathcal{E}^c|$ by choosing $\gamma_{n,jj'} = \frac{\pi}{2}\sdtauj\Phi^{-1}(1-\frac{f}{p(p-1)})/\sqrt{n}$, where 
	\begin{equation} \label{sigmaij}
		\vartauj=\frac{4(n-1)}{(n-2)^2}\sum_{i'=1}^{n}\left[ \left\lbrace \frac{1}{n-1}\sum_{i=1, i\neq i'}^{n} \Sign \left(\left(X_{ij}-X_{i'j}\right)\left(X_{ij'}-X_{i'j'}\right)\right)\right\rbrace -\widehat{\tau}_{jj'}\right]^{2},
	\end{equation}
	and $\widehat{\tau}_{jj'}$ is defined in \eqref{kendalltau}. 
Furthermore, with this threshold, the sure screening property of Theorem~\ref{control_fnr} still holds.
\end{theorem}
The estimator \eqref{sigmaij}  is a jackknife estimator for the asymptotic variance of a U-statistic \citep{ar69,sen77, cv81, fr83, lee85}.

\subsection{A Second Look at Assumptions~\ref{min_rho} and \ref{max_rho}} \label{subsec:prop}
Assumptions~\ref{min_rho} and \ref{max_rho}  involve placing conditions  on the elements of $\Sigma$ corresponding to non-zero and zero elements of $\Sigma^{-1}$, respectively. These conditions are somewhat hard to interpret, since in general there is no simple relationship between the $(j,j')$th elements of a matrix $A$ and its inverse $A^{-1}$. We now present a result  from \citet{luo12} that allows us to re-formulate these assumptions as conditions on the elements of $\Sigma^{-1}$. 
We let $\beta=\Lambda_{\max}(\Sigma)/\Lambda_{\min}(\Sigma)$ and $\nu={2}/\left\{ \Lambda_{\max}(\Sigma)^{-1}+\Lambda_{\min}(\Sigma)^{-1} \right\}$. 
 Here, $\Lambda_{\max}(\cdot)$ and $\Lambda_{\min}(\cdot)$ indicate the largest and smallest eigenvalues of a matrix, respectively. 

\begin{proposition} \citep{luo12} \label{prop3}
	Suppose $1 < \beta \leq \left\{n^{(1-\xi)/2}+\Lambda_{\max}(\Sigma)^{-1/2}  \right\}/\left\{n^{(1-\xi)/2}-\Lambda_{\max}(\Sigma)^{-1/2}  \right\}$.
	Then   Assumption~\ref{max_rho} holds. Suppose also that $n \geq \left(2/C_1\right)^{1/(1-\xi-\kappa)}$.
	If $\min_{(j,j') \in \mathcal{E}} \nu^2 \left| (\Sigma^{-1})_{jj'} \right| \geq 2C_1 n^{-\kappa}$, then Assumption~\ref{min_rho} holds. Furthermore, if Assumption~\ref{min_rho} holds, then $\min_{(j,j') \in \mathcal{E}} \nu^2 \left| (\Sigma^{-1})_{jj'} \right| \geq C_1 n^{-\kappa}/2$.
\end{proposition}
  Therefore, if $\Sigma$ is very well-conditioned and the non-zero elements of $\Sigma^{-1}$ are sufficiently large, then Proposition~\ref{prop3} implies that Theorems~\ref{control_fnr} and \ref{control_fpr} hold.


\section{Simulation Studies}\label{sec:sim}

\indent \indent The simulation studies in this section are largely based on those in \citet{luo12}.

\subsection{Data Generation} \label{sec:datagen}
\indent\indent Let $p$ be the number of features, and $n$  the number of observations. Motivated by the simulation study of \citet{luo12}, we considered four ways of generating the edge set $\mathcal{E}$.
\begin{list}{}{}
	\item[\textbf{\emph{Simulation A:}}] For all $j<j'$, we set $(j,j') \in \mathcal{E}$ with probability $0.01$. We then generated a $p \times p$ matrix $A$, where
	\begin{equation} \label{A} 
	 A_{jj'} = A_{j'j} = \begin{cases}
	1 & \text{for $j=j'$}\\ \mathrm{Unif}[-0.3,0.7] & \text{for $(j,j') \in \mathcal{E}$ } \\ 0 & \text{otherwise} \end{cases}. 
	\end{equation} 
	 
	Finally, we created a positive definite matrix ${\Sigma}^{-1}$,
	\begin{equation} \label{sigmainv}
	{\Sigma}^{-1} = {A} + (0.1 - \Lambda_{\min}({A})){I},
	\end{equation}
	  
	where $\Lambda_{\min}(A)$ is the smallest eigenvalue of $A$, and $I$ denotes the $p \times p$ identity matrix.

	\item[\textbf{\emph{Simulation B:}}]
	We partitioned the $p$ features into $10$ equally-sized and non-overlapping sets, $C_l=\{(l-1)p/10+1, \ldots, lp/10\}$  for $l=1, \ldots, 10$.  
	For all $j \in C_l, j' \in C_l, j<j'$, we set $(j,j') \in \mathcal{E}$. We then generated $A$ and $\Sigma^{-1}$ according to \eqref{A} and \eqref{sigmainv}. 
	
	\item[\textbf{\emph{Simulation C: }}]
	 For all $j\leq j'$, we set $\rho_{jj'}=0.3^{|j-j'|}$. 
	
	\item[\textbf{\emph{Simulation D: }}]  
	We partitioned the features into $p/10$ equally-sized and non-overlapping sets, $C_l= \{10(l-1)+1,\ldots,10l \} $ for $l=1,\ldots, p/10$. Then for all $j \in C_l, j' \in C_l$, we set $(\Sigma^{-1})_{jj'}=0.9^{|j-j'|}$. All other elements of $\Sigma^{-1}$ were set to zero.  
	
\end{list}

$\Sigma$ was rescaled to have diagonal elements equal to 1. We then generated observations from a $N(0, {\Sigma})$ distribution, and observations from a multivariate $t$-distribution with $\theta$ degrees of freedom, mean zero, and correlation $\Sigma$. After that, we applied four monotonic functions, $\exp(x)$, $x^{3}$, $x^{5}$, $(x-1)^{3}$, to these observations with equal probability; this process gave us nonparanormal-distributed observations and transelliptical $t$-distributed observations.

\subsection{Control of False Positive Rate}

Theorem~\ref{control_fpr} states that under certain conditions, performing transelliptical GRASS with $\gamma_{n,jj'} = \frac{\pi}{2}\sdtauj\Phi^{-1}(1-\frac{f}{p(p-1)})/\sqrt{n}$, where $\sdtauj$ is of the form \eqref{sigmaij}, leads to control of the asymptotic expected false positive rate (FPR) at level  $q \equiv f/|\mathcal{E}^c|$.  
 In Tables~\ref{tab:fprfnr1} and \ref{tab:fprfnrt2}, we explore the control of the FPR in finite samples, for nonparanormal and transelliptical $t$-distributed data.
  The FPR (defined as FP/(FP+TN)) and false negative rate (FNR; defined as FN/(TP+FN)) are reported for various values of the level of desired FPR control, $q$. The size of the estimated edge set $|\ehg|$ is also reported. Here $n=100$ and $p=1000$, and results are averaged over 250 simulated data sets. 
  


Assumption~\ref{max_rho} is the key for controlling the asymptotic expected false positive rate. In Simulation B, both $\Sigma$ and ${\Sigma}^{-1}$ are block diagonal with ten completely dense blocks. So Assumption~\ref{max_rho} holds in Simulation B. In Simulation D, all of the zero elements in the precision matrix ${\Sigma}^{-1}$ also correspond to zero elements in $\Sigma$, so that Assumption~\ref{max_rho} holds exactly. As expected, the FPR is controlled successfully in Simulations B and D.

In contrast, in Simulations A and C, not all of the zero elements in the precision matrix ${\Sigma}^{-1}$ correspond to zero elements in the correlation matrix $\Sigma$. But Table~\ref{tab:fprfnr1} and Table~\ref{tab:fprfnrt2} reveal that the FPR is still controlled well in these settings. This is because Assumption~\ref{max_rho} only requires the elements in $\mathcal{E}^c$ to correspond to small, though not necessarily zero, elements of $\Sigma$. This assumption holds for most of the elements in $\mathcal{E}^c$ and $\mathcal{E}$ for Simulations A and C. Therefore, the FPR is also well-controlled in Simulations  A and C.

\begin{table}[h!]
  \centering
   \caption{False positive rate control for nonparanormal data using $\gamma_{n,jj'} = \frac{\pi}{2}\sdtauj\Phi^{-1}(1-\frac{f}{p(p-1)})/\sqrt{n}$, where $\sdtauj$ is of the form \eqref{sigmaij} in Theorem~\ref{control_fpr}. \label{tab:fprfnr1}}
  \begin{adjustbox}{width=1\textwidth}

		\begin{tabular}{|c|ccc|ccc|ccc|ccc|}
			\hline
			&   \multicolumn{3}{|c|}{Simulation A}& \multicolumn{3}{|c|}{Simulation B}& \multicolumn{3}{|c|}{Simulation C}& \multicolumn{3}{|c|}{Simulation D}\\
				$q$	 &$|\ehg|$	 &FPR	 &FNR	 &$|\ehg|$	 &FPR	 &FNR	&$|\ehg|$	 &FPR	 &FNR &$|\ehg|$	 &FPR	 &FNR \\
			\hline
			1e-04   &542.46	 &3e-04	         &0.925  &1611.6	 &2e-04	         &0.969  &290.22	 &2e-04	         &0.819  &920.38	 &2e-04         &0.819	 \\
			0.001  &1579.76	 &0.0019   &0.868  &3472.04	&0.0014   &0.943  &1071.96	 &0.0014  	 &0.648 &1580.36	 &0.0014	         &0.806 \\
			0.01   &7426.36	 &0.013	 &0.762  &10950.86	 &0.011	 &0.879  &6125.34	 &0.011	 &0.382  &6364.42	 &0.011	         &0.793 \\
			0.1    &53548.1	 &0.104	 &0.551  &59265.12	 &0.098	 &0.695 &49926.3	 &0.098	 &0.111 &49884.46	 &0.098	 &0.722\\
			0.2    &102830.74	 &0.202	 &0.446  &108698.82	 &0.195	 &0.580  &98455.7	 &0.196	 &0.055 &98333.9	 &0.195	 &0.644\\
			0.3    &152029.24	 &0.301	 &0.368  &157545.2	 &0.294	   &0.488  &147546.26	 &0.294	 &0.031 &147393.94	 &0.294	 &0.566 \\
			0.5    &250794.72	 &0.500	 &0.244  &254988.28	 &0.493	 &0.332    &246989.82	 &0.493	 &0.012 &246737.66	 &0.493	 &0.405 \\
			\hline
			\end{tabular}
	\end{adjustbox}

\end{table}

\begin{table}[h!]
	\centering
	\caption{False positive rate control for transelliptical $t$-distributed data with 5 degrees of freedom using  $\gamma_{n,jj'} = \frac{\pi}{2}\sdtauj\Phi^{-1}(1-\frac{f}{p(p-1)})/\sqrt{n}$, where $\sdtauj$ is of the form \eqref{sigmaij} in Theorem~\ref{control_fpr}. \label{tab:fprfnrt2}}
				\begin{adjustbox}{width=1\textwidth}
						\begin{tabular}{|c|ccc|ccc|ccc|ccc|}
						\hline
						&   \multicolumn{3}{|c|}{Simulation A}& \multicolumn{3}{|c|}{Simulation B}& \multicolumn{3}{|c|}{Simulation C}& \multicolumn{3}{|c|}{Simulation D}\\
							$q$	 &$|\ehg|$	 &FPR	 &FNR	 &$|\ehg|$	 &FPR	 &FNR	&$|\ehg|$	 &FPR	 &FNR &$|\ehg|$	 &FPR	 &FNR \\
						\hline
					1e-04  &448.836	 &3e-04	         &0.941  &1325.172	 &2e-04	         &0.975  	&240.52	 &2e-04	         &0.87  &859.488	 &2e-04	         &0.833 \\
					0.001  &1447.58	 &0.0018   &0.889  &3039.336	&0.0015   &0.952   &1012.12	 &0.0015  	 &0.721 &1564.668	 &0.0015	         &0.813 \\
					0.01   &7261.952	 &0.013	 &0.788  &10340.736	 &0.011	 &0.892  &6123.648	 &0.0112	 &0.466  &6438.42	 &0.0111	         &0.794\\
					0.1    &53507.624	 &0.104	 &0.575  &58758.12	 &0.099	 &0.713  &50330.904	 &0.099	 &0.158 &50297.052	 &0.099	 &0.721 \\
					0.2    &102902.456	 &0.203	 &0.468  &108367.908	 &0.197	 &0.598  &99069.432	 &0.197	 &0.085 &98909.556	 &0.197	 &0.642  \\
					0.3    &152190.388	 &0.302	 &0.387  &157358.224	 &0.295	   &0.504  &148208.564	 &0.295	 &0.053 &148021.74	 &0.295	 &0.563 \\
					0.5    &250987.548	 &0.50	 &0.258   &254984.984	 &0.494	 &0.344  &247579.316	 &0.495	 &0.023 &247400.096	 &0.494	 &0.404  \\
					\hline
				\end{tabular}
		\end{adjustbox}
		
	\end{table}

\subsection{Comparison to Existing Approaches}

\begin{figure}[h!]
	\centering
	\includegraphics[width=0.88\textwidth]{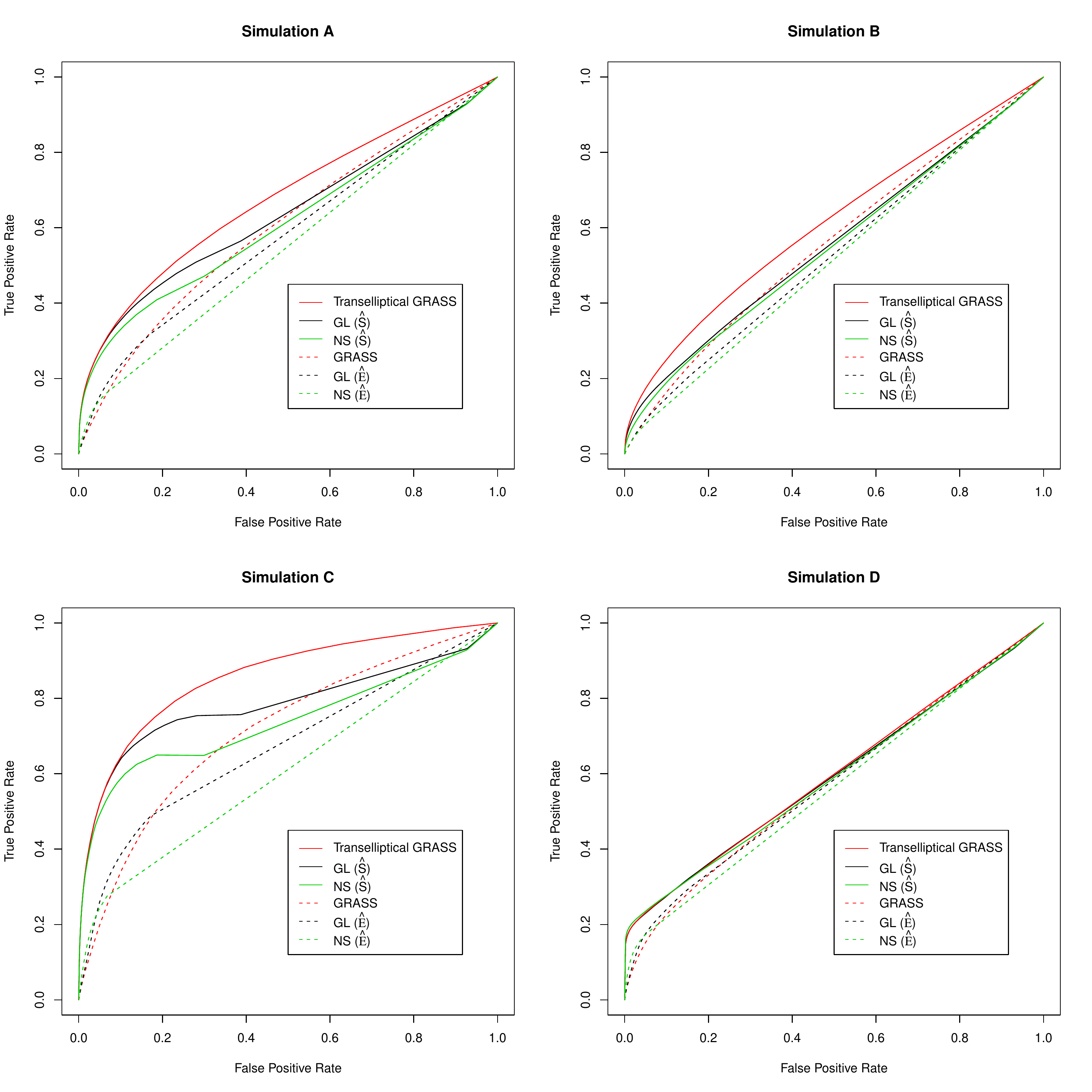}
	\caption{   \label{results:abc} For nonparanormally distributed data with $p=750$ and $n=50$, the true positive and false positive rates are shown. Curves are obtained by varying the tuning parameter for each method, and results are averaged over 20 simulated data sets. }
\end{figure}

\indent \indent We first compare the performances of the graphical lasso \citep{fr08}, neighborhood selection \citep{mb06}, and transelliptical GRASS on nonparanormal data generated from Simulations A--D, with $n=50$ and $p=750$. Let $X$ denote the $n \times p$ simulated data matrix. We let GL($\widehat{\Sigma}$) and NS($\widehat{\Sigma}$) denote the results of the graphical lasso and the neighborhood selection applied to the estimated correlation matrix $\widehat{\Sigma}=X^{T}X/n$; and we let GL($\widehat{S}^{\tau}$) and NS($\widehat{S}^{\tau}$) denote the results of the graphical lasso and neighborhood selection applied to $\widehat{S}^{\tau}$, the Kendall's tau estimator, defined in \eqref{stau}. 
  Results are shown in Figure~\ref{results:abc}.

Transelliptical GRASS outperforms GL($\widehat{S}^{\tau}$) and NS($\widehat{S}^{\tau}$) in Simulation B. The sparsity patterns of $\Sigma$ and ${\Sigma}^{-1}$ are identical, so the assumptions underlying transelliptical GRASS hold.  

In Simulations A and C, most of the edges in $\mathcal{E}$ correspond to large elements of $\Sigma$, and most of the non-edges in $\mathcal{E}$ correspond to small elements of $\Sigma$. Consequently,  transelliptical GRASS outperforms GL($\widehat{S}^{\tau}$) and NS($\widehat{S}^{\tau}$).

Simulation D was designed to violate Assumption~\ref{min_rho}; most of the elements in the edge set $\mathcal{E}$ correspond to zero values in the correlation matrix $\Sigma$. Therefore transelliptical GRASS does not perform well in Simulation D. But even in this undesirable setting, Figure~\ref{results:abc} indicates that GL($\widehat{S}^{\tau}$) and NS($\widehat{S}^{\tau}$) do not perform much better than the transelliptical GRASS.

Not surprisingly, the original GRASS proposal (which involves thresholding $\widehat{\Sigma}$),  GL($\widehat{\Sigma}$), and NS($\widehat{\Sigma}$) perform poorly, because they are designed for Gaussian data rather than nonparanormal data.

\begin{figure}[h!]
	\centering
	\includegraphics[width=0.88\textwidth]{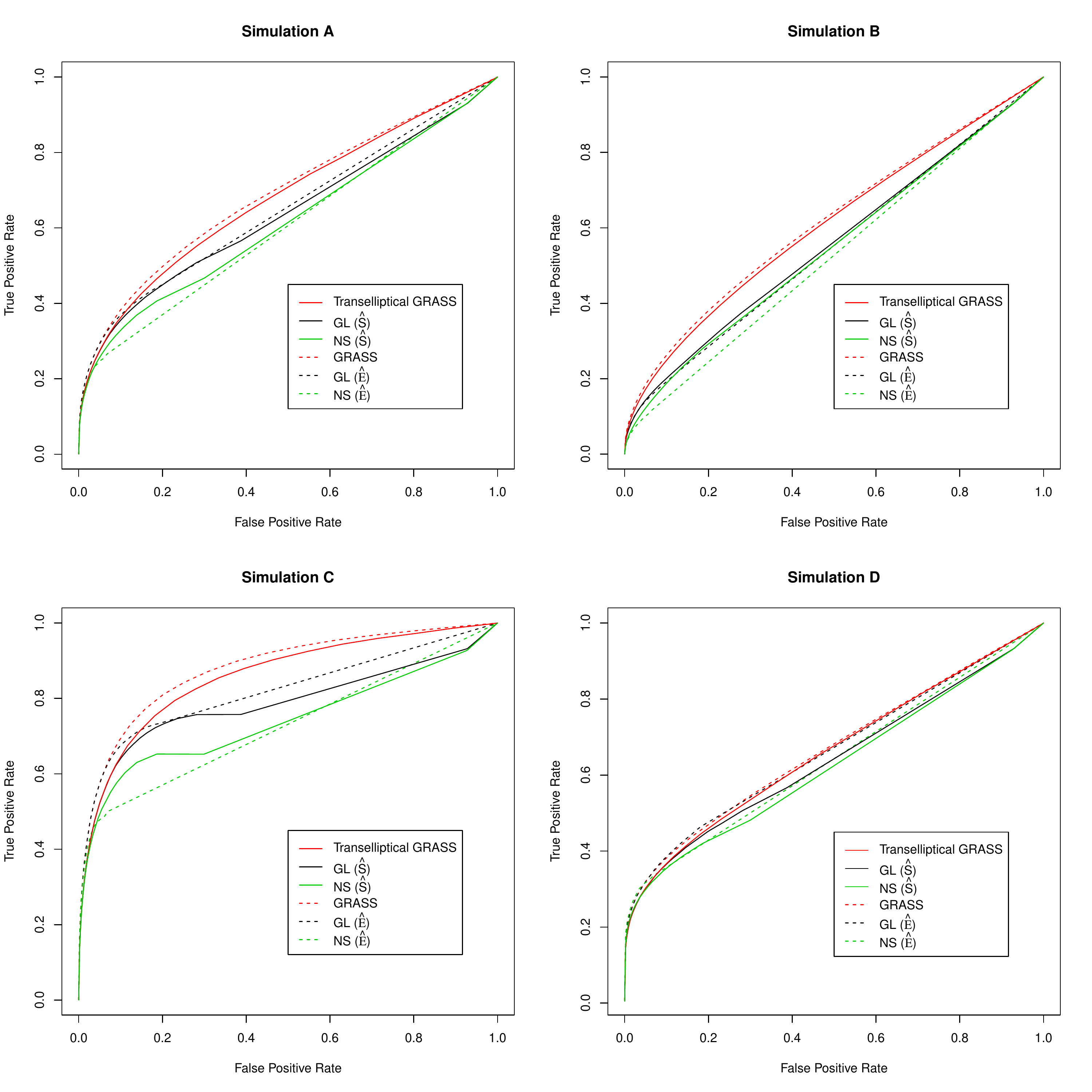}
	\caption{   \label{results:normal} For Gaussian-distributed data with $p=750$ and $n=50$, the true positive and false positive rates are shown. Curves are obtained by varying the tuning parameter for each method, and results are averaged over 20 simulated data sets. }
\end{figure}

Finally, we compare the performance of the aforementioned methods on Gaussian data generated from Simulations A--D, with $n=50$ and $p=750$. Figure~\ref{results:normal} shows that the original GRASS proposal performs only slightly better than transelliptical GRASS on Gaussian data. This suggests that when the underlying distribution of the data is unknown, there is little cost (and potentially a large gain) associated with performing transelliptical GRASS instead of GRASS.

Overall, Figures~\ref{results:abc} and \ref{results:normal} suggest that in these four settings, transelliptical GRASS performs competitively compared to some popular but computationally-intensive procedures for estimating a precision matrix.

\section{Application to Equities data} \label{equities}
We examined the Yahoo! Finance stock price data, which is described in \citet{liu12b}, and available in the \verb=huge= package in \verb=R= on CRAN. The data consists of 1258 daily closing prices for 452 stocks in the S$\&$P 500 index between January 1, 2003 and January 1, 2008. The stocks are categorized into 10 Global Industry Classification Standard (GICS) sectors.
Let $S_{ij}$ denote the closing price of the $j$th stock on the $i$th day. We construct
a $1257 \times 452$ data matrix $X$ such that $X_{ij} = \log\left(S_{(i+1)j}/S_{ij}\right)$ for $i = 1, \dots , 1257$ and $j = 1, \dots , 452$. We standardize each stock to have mean zero and standard deviation one, as in \citet{km15}.

\begin{figure}[h!]\
	\includegraphics[width=0.51\textwidth]{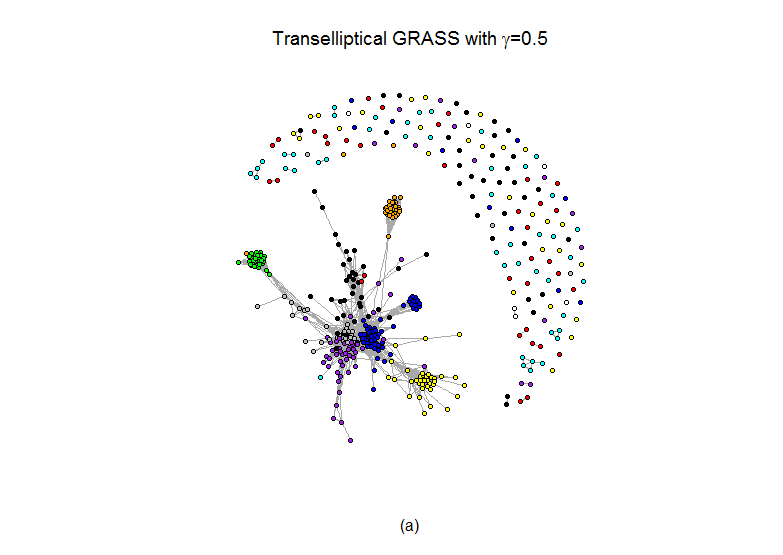}\hfill \includegraphics[width=0.51\textwidth]{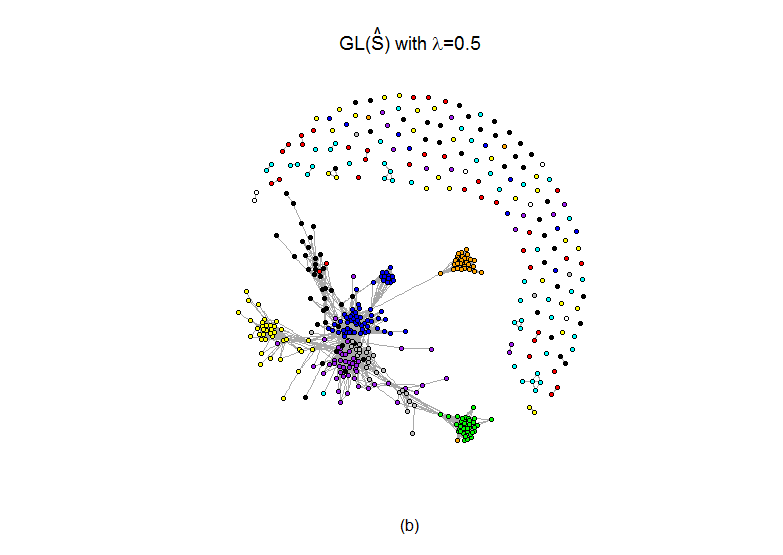}\hfill
	\includegraphics[width=0.51\textwidth]{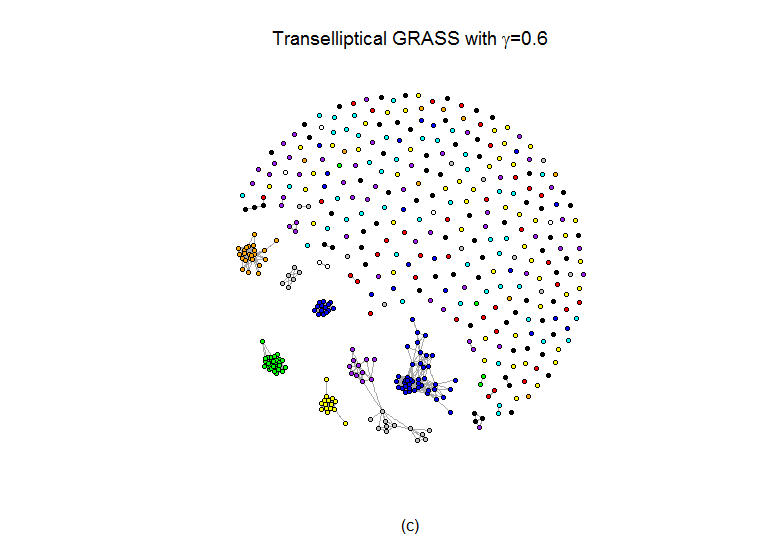}\hfill
	\includegraphics[width=0.51\textwidth]{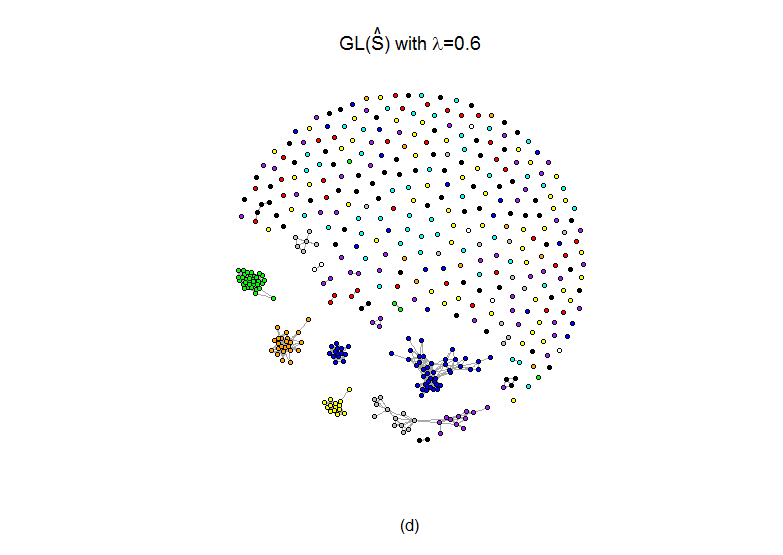}\hfill
	\includegraphics[width=0.51\textwidth]{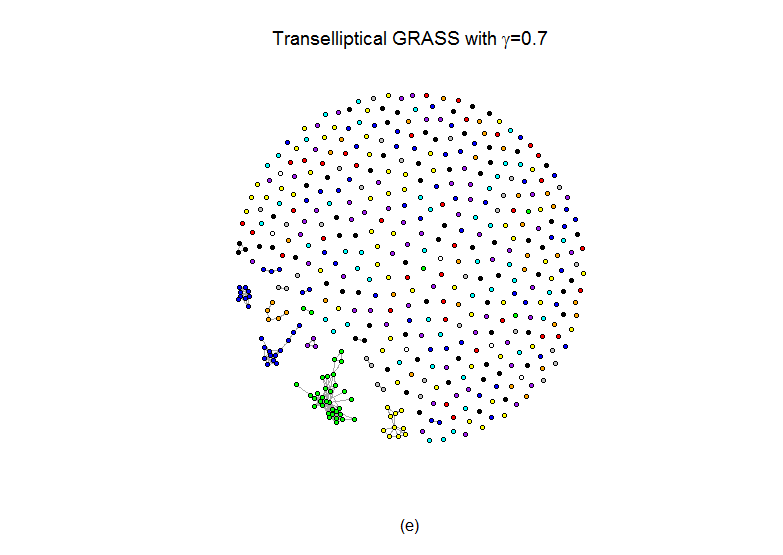}\hfill
	\includegraphics[width=0.51\textwidth]{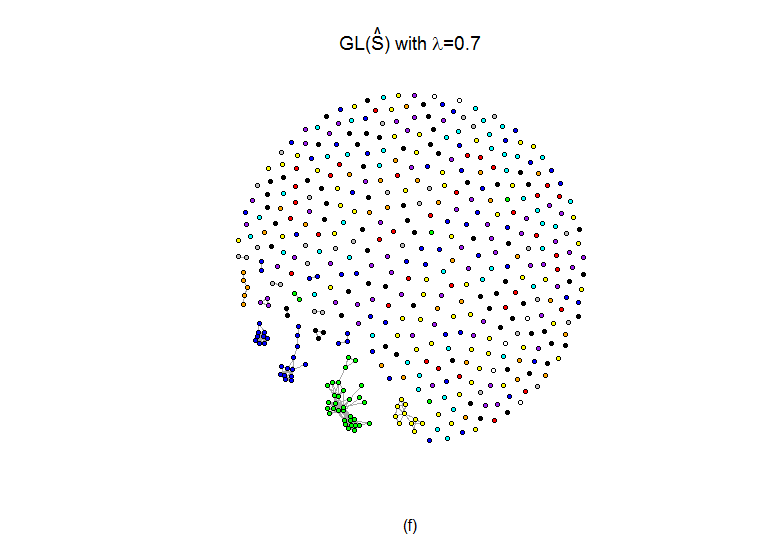}		
	\caption{   \label{realdata}  $(a)$: Transelliptical GRASS with $\gamma_{n,jj'}=0.5$ (3336 edges). $(b)$: GL($\widehat{S}^{\tau}$) with $\lambda = 0.5$ (2343 edges). $(c)$: Transelliptical GRASS with $\gamma_{n,jj'}=0.6$ (1036 edges). $(d)$: GL($\widehat{S}^{\tau}$) with $\lambda = 0.6$ (847 edges). $(e)$: Transelliptical GRASS with $\gamma_{n,jj'}=0.7$ (232 edges). $(f)$: GL($\widehat{S}^{\tau}$) with $\lambda = 0.7$ (222 edges). The nodes are colored based on the GICS sector of the corresponding stock.}
\end{figure}

We applied transelliptical GRASS with $\gamma_{n,jj'}$=0.5, 0.6, 0.7, and GL($\widehat{S}^{\tau}$) with $\lambda$=0.5, 0.6, 0.7. 
 Figure~\ref{realdata} indicates that in all estimated graphs, stocks from the same GICS sector tend to be highly-connected, indicating that both methods provide informative graph estimates. Furthermore, plots $(c)$, $(d)$, $(e)$ and $(f)$ of Figure~\ref{realdata} also show that the graph estimates from transelliptical grass and GL($\widehat{S}^{\tau}$) are quite similar provided that  $\gamma_{n,jj'}=\lambda$. In fact, the arguments in \citet{witten2011new} can be used to establish the fact that when $\gamma_{n,jj'}=\lambda$, the connected components of GRASS and GL($\widehat{S}^{\tau}$) are identical.

\section{Discussion} \label{discussion}
In this paper, we have proposed transelliptical GRASS, a simple and efficient procedure for recovering the structure of a high-dimensional transelliptical graphical model. Transelliptical GRASS is a natural extension of the GRASS proposal of \cite{luo12} to the non-normal setting. Transelliptical GRASS shares the attractive theoretical and computational properties of the original GRASS proposal. We have established that it performs almost as well as methods that assume Gaussianity when the data are Gaussian, and much better than methods that assume Gaussianity in the case of non-Gaussian data.  Therefore, in general, there is little cost to applying transelliptical GRASS instead of the original GRASS proposal.




\appendix
\section{Appendix}
\subsection{Proof of Theorem~\ref{control_fnr}}

\begin{definition} \label{one_sample_U}
	 (\citet{hoef63}, page 25) Let $W_1, W_2, \ldots, W_n$ be independent random variables. For $m \leq n$, a one-sample U-statistic takes the form 
	 \begin{equation} \label{U_stat}
	 U = \frac{1}{n^{(m)}} \sum_{n,m} g\left(W_{i_1}, \ldots, W_{i_m}\right),
	 \end{equation}
	 where $n^{(m)}=n \left(n-1\right) \ldots \left(n-m+1\right)$, and the sum $\sum_{n,m}$ is taken over all $m$-tuples $i_1, \ldots, i_m$ of distinct positive integers not exceeding $n$. 
\end{definition}

\begin{lemma} \label{hoef63}
	 (\citet{hoef63}, page 25) If the function of g in \eqref{U_stat} is bounded
	as $a\leq g\left(W_{i_1}, \ldots, W_{i_m}\right)\leq b$, then for any $t>0$ and $m\leq n$, we have
	\begin{equation*}
	\Pr\left\{\left|U-\Mean\left(U\right)\right|>t\right\}\leq 2\exp\left\{\frac{-2\lfloor n/m\rfloor t^2}{(b-a)^2}\right\}.
	\end{equation*}
\end{lemma}


	 
Recall from \eqref{kendalltau} that $$\widehat{\tau}_{jj'}=\frac{2}{n(n-1)}\sum_{1\leq i\leq i'\leq n}\Sign \left( \left(X_{ij}-X_{i'j}\right)\left(X_{ij'}-X_{i'j'}\right)\right).$$ Let $W_i=(X_{ij}, X_{ij'})$, $i=1, \ldots, n$. Then the sample estimator $\widehat{\tau}_{jj'}$ is of the form \eqref{U_stat} with $m=2$ and $g\left(W_i,W_{i'}\right)=\Sign\left(\left(X_{ij}-X_{i'j}\right)\left(X_{ij'}-X_{i'j'}\right)\right)$.
Therefore, we can apply Lemma \ref{hoef63}  with $m=2$ and $-1 \leq g\left(W_i,W_{i'}\right) \leq 1$,  which yields 
\begin{equation*}
\Pr\left(\left|\widehat{\tau}_{jj'}-\tau_{jj'}\right|>\frac{2}{3\pi}C_1 n^{-\kappa}\right)\leq 2 \exp \left( -\frac{2\lfloor n/2 \rfloor}{9\pi^{2}}C_1^2 n^{-2\kappa} \right).
\end{equation*}



Next, we notice that
\begin{eqnarray} 
\Pr\left(\left|\widehat{S}^{\tau}_{jj'}-\rho_{jj'}\right|>\frac{1}{3}C_1 n^{-\kappa}\right)&=& 
\Pr\left(\left|\sin\left(\frac{\pi}{2}\widehat{\tau}_{jj'}\right)-\sin\left(\frac{\pi}{2}\tau_{jj'}\right)\right|>\frac{1}{3}C_1 n^{-\kappa}\right) \nonumber \\
&\leq&\Pr\left(\left|\widehat{\tau}_{jj'}-\tau_{jj'}\right|>\frac{2}{3\pi}C_1 n^{-\kappa}\right) \nonumber \\
&\leq& 2\exp\left( -\frac{2\lfloor n/2 \rfloor}{9\pi^{2}}C_1^2 n^{-2\kappa}\right).\label{concineq}
\end{eqnarray}
The first equality results from directly applying Lemma \ref{lemmaliu} and our definition of $\widehat{S}^{\tau}_{jj'}$. The first inequality results from applying the mean value theorem. 
  It follows that 
\begin{eqnarray*}
\Pr\left(\mathcal{E} \nsubseteq\ehg\right)&=&\Pr\left\{\bigcup_{(j,j')\in\mathcal{E}} \left( \left|\widehat{S}^{\tau}_{jj'}\right|<\gamma_n \right)\right\}\\
& \leq &\sum_{(j,j')\in\mathcal{E}}\Pr\left(\left|\widehat{S}^{\tau}_{jj'}\right| <\frac{2}{3}C_1n^{-\kappa}\right)\\
&\leq& \sum_{(j,j')\in\mathcal{E}} \Pr\left(\left|\widehat{S}^{\tau}_{jj'}-\rho_{jj'}\right|\geq\frac{1}{3}C_1 n^{-\kappa}\right)\\
&\leq& 2p^2 \exp\left( -\frac{2\lfloor n/2 \rfloor}{9\pi^{2}}C_1^2 n^{-2\kappa}\right).
\end{eqnarray*}
Here, the second inequality follows from Assumption~\ref{min_rho}, and the third inequality from the fact that $\mid \mathcal{E} \mid \leq p^2$ along with \eqref{concineq}.


Therefore, we have shown that $$\Pr(\mathcal{E}\subseteq\ehg) \geq 1-2p^2\cdot\exp\left(-\frac{2\lfloor n/2 \rfloor}{9\pi^{2}}C_1^2 n^{-2\kappa}\right).$$ A similar argument can be used to establish that $$\Pr\left(\mathcal{E}_j\subseteq\ehag\right) \geq 1-2p^2\cdot\exp\left(-\frac{2\lfloor n/2 \rfloor}{9\pi^{2}}C_1^2 n^{-2\kappa}\right).$$

Conversely, if
$\min_{(j,j')\in \mathcal{E}}|\rho_{jj'}|< C_1n^{-\kappa}/3$,
then there exists $(j^{\star},{j'}^{\star})\in \mathcal{E}$ such that $|\rho_{j^{\star}{j'}^{\star}}|<C_1n^{-\kappa}/3$. This together with \eqref{concineq}
implies that
\begin{eqnarray*}
	\Pr(\mathcal{E} \subseteq \ehg) \leq \Pr\left\{(j^{\star},{j'}^{\star}) \in \ehg\right\}\le \Pr\left(\left| \widehat{S}^{\tau}_{j^{\star}{j'}^{\star}}-\rho_{j^{\star}{j'}^{\star}} \right| \geq C_1n^{-\kappa} /3 \right)\leq 2\exp\left( -\frac{2\lfloor n/2 \rfloor}{9\pi^{2}}C_1^2 n^{-2\kappa}\right),
\end{eqnarray*}
so that the result holds.

\subsection{Proof of Corollary~\ref{concomp}}
\begin{proof}
	It suffices to show that transelliptical GRASS will not result in edges between $x_{j}^{(s)}$ and $x_{j'}^{(t)}$ for all $s\not=t$ with high probability. 
	
	This is  the case 
	when the event $\left(\left|\widehat{S}^{\tau}_{jj'}-\rho_{jj'}\right|\leq\frac{1}{3}C_1 n^{-\kappa}\right)$ holds for all $j \neq j'$. 
	As was shown in the proof of Theorem~\ref{control_fnr}, 
	$\Pr\left\{ \bigcap_{j \neq j'} \left|\widehat{S}^{\tau}_{jj'}-\rho_{jj'}\right|\leq\frac{1}{3}C_1 n^{-\kappa} \right\} \geq 1 - C_4\mathrm{exp}\left(-C_5n^{1-2\kappa}\right)$. 
\end{proof}

\subsection{Proof of Theorem~\ref{control_size}}

Let
\begin{equation*}
	L_j=\left\{j':j'\neq j,\left|\rho_{jj'}\right|\geq\frac{1}{3}C_1n^{-\kappa}\right\}
\end{equation*}
and
\begin{equation*}
	\Gamma_{j}=\bigcap_{j':j'\neq j}\left\{\left|\widehat{S}^{\tau}_{jj'}-\rho_{jj'}\right|\leq \frac{1}{3}C_1n^{-\kappa}\right\}.
\end{equation*}
 By definition, $\ehag=\left\{j':j'\neq j,\left|\widehat{S}^{\tau}_{jj'}\right|>\frac{2}{3}C_1n^{-\kappa}\right\}$. On the set $\Gamma_{j}$, if $j'$ belongs to $\ehag$, it has to belong to $L_j$. Thus, we conclude that $\Pr\left(\ehag\subseteq L_j\right) \geq \Pr\left(\Gamma_{j}\right)$.  An argument similar to that in the proof of Theorem~\ref{control_fnr} can be used to show that 
\begin{equation*}
	\Pr\left(\Gamma_{j}\right)\geq 1-C_4\exp\left(-C_5n^{1-2\kappa}\right).
\end{equation*}
This implies that 
\begin{equation}
	\Pr\left(\ehag\subseteq L_j\right)\geq 1-C_4\exp\left(-C_5n^{1-2\kappa}\right).
	\label{ELineq}
\end{equation}

Define $D=\sum_{j' \in L_j}\rho_{jj'}^2$. Then, by the definition of $L_j$, it follows that 
\begin{equation}
D \ge \frac{1}{9}C_1^2\left|L_j\right| n^{-2\kappa}.
\label{Alowerbound}
\end{equation}
Furthermore, 
\begin{equation}
D\le \sum_{j'=1}^p \rho_{jj'}^2= \left\|\Sigma e_j\right\|_2^2 \le \Lambda_{\max}\left(\Sigma\right)e_j^T \Sigma e_j=\Lambda_{\max}\left(\Sigma\right),
\label{Aupperbound}
\end{equation}
where $e_j$ is the unit vector with a one in the $j$th element and zeros elsewhere, and where the last equality results from the fact that the diagonal elements of $\Sigma$ are equal to $1$.

Combining \eqref{Alowerbound} and \eqref{Aupperbound}, we have that
$$|L_j| \leq  9C_1^{-2}n^{2\kappa}\Lambda_{\max}\left(\Sigma\right).$$ This, in conjunction with Assumption~\ref{max_eigen} and \eqref{ELineq}, completes the proof of Theorem~\ref{control_size}.

\noindent

\subsection{Proof of Theorem~\ref{control_fpr}}
First, we verify the sure screening property (Section~\ref{subsec:surescreen}). We then establish the control of the asymptotic expected false positive rate (Section~\ref{subsec:controlfpr}).
\subsubsection{Verification of Sure Screening Property}\label{subsec:surescreen}
To show that the sure screening property holds, it is enough to show that
\begin{equation}
\gamma_{n,jj'}=\frac{\pi}{2}\sdtauj\Phi^{-1}\left(1-\frac{f}{p(p-1)}\right)/\sqrt{n}\leq\frac{2}{3}C_1n^{-\kappa}.
\end{equation}
 In other words, we must show that
 \begin{equation}
 	\frac{f}{p(p-1)}  \geq 1 - \Phi\left( \frac{4}{3 \pi \sdtauj}C_1n^{\frac{1}{2}-\kappa} \right) .
 	\label{enough2}
 \end{equation}
 Recall that $1-\Phi\left(x\right)\leq\frac{1}{\sqrt{2\pi}}x^{-1}\exp\left(-x^2/2\right)$, which implies that
 \begin{equation} \label{64}
 	1-\Phi\left(\frac{4}{3\pi\sdtauj}C_1n^{\frac{1}{2}-\kappa}\right) \leq C_8 n^{-\frac{1}{2}+\kappa}\exp\left(-C_9 n^{1-2\kappa}\right).
 \end{equation}
 Since $\log\left(p\right)=C_3n^{\xi}$, we have that 
 \begin{equation}
 	f/\left\{p(p-1)\right\}\geq C_{10}\exp\left(-C_{11}n^{\xi}\right).
 	\label{fpp1}
 \end{equation} 
 Combining \eqref{64} and \eqref{fpp1}, and using the fact that $\xi<1-2\kappa$, 
 \eqref{enough2} follows directly.

\subsubsection{Control of the Asymptotic Expected False Positive Rate} \label{subsec:controlfpr}
Next, we show that the choice of  $\gamma_{n,jj'}$ given in the statement of Theorem~\ref{control_fpr} leads to  control of the asymptotic expected false positive rate at $f/|\mathcal{E}^c|$. The following lemma is used here.

\begin{lemma} \label{arv69}
	Consider two random variables $X_j$ and $X_{j'}$, each with $n$ i.i.d observations. Then
	\begin{equation}
	\frac{\sqrt{n}(\widehat{\tau}_{jj'}-\tau_{jj'})}{\sdtauj}\stackrel{d}\longrightarrow N(0,1), 
	\end{equation}
	where $\widehat{\tau}_{jj'}$ is defined in \eqref{kendalltau} and $\vartauj$ is defined in \eqref{sigmaij}. 
\end{lemma}
Lemma~\ref{arv69} follows from Theorem 6 in \cite{ar69} in conjunction with Slutsky's Theorem. It follows from an application of the delta method that
\begin{equation} \label{lemma5}
\frac{\sqrt{n}\left(\widehat{S}_{jj'}^{\tau}-\rho_{jj'}\right)}{\frac{\pi}{2}\sdtauj\sqrt{\left(1-\rho^2_{jj'}\right)}} \stackrel{d}\longrightarrow N(0,1).
\end{equation} 

Therefore, for any $(j,j')\notin\mathcal{E}$, we have

\begin{align*}
\Pr\left(\left|\widehat{S}_{jj'}^{\tau}\right|>\gamma_{n,jj'}\right) &= \Pr\left(\frac{\sqrt{n}\left(\widehat{S}_{jj'}^{\tau}-\rho_{jj'}\right)}{\frac{\pi\sdtauj}{2}\sqrt{1-\rho^2_{jj'}}}>\frac{\sqrt{n}\left(\gamma_{n,jj'}-\rho_{jj'}\right)}{\frac{\pi\sdtauj}{2}\sqrt{1-\rho^2_{jj'}}}\right)\\ 
&\ \ \ +\Pr\left(\frac{\sqrt{n}\left(\widehat{S}_{jj'}^{\tau}-\rho_{jj'}\right)}{\frac{\pi\sdtauj}{2}\sqrt{1-\rho^2_{jj'}}}<-\frac{\sqrt{n}\left(\gamma_{n,jj'}+\rho_{jj'}\right)}{\frac{\pi\sdtauj}{2}\sqrt{1-\rho^2_{jj'}}}\right)\\
&\to 1-\Phi\left(\Phi^{-1}\left(1-\frac{f}{p(p-1)}\right)\right)+1-\Phi\left(\Phi^{-1}\left(1-\frac{f}{p(p-1)}\right)\right)\\
& =\frac{2f}{p(p-1)},
\end{align*}
where the convergence results from combining \eqref{lemma5}, Assumption~\ref{max_rho}, and the fact that the order of $\sqrt{n}\gamma_{n,jj'}$ is much larger than that of $\sqrt{n}\rho_{jj'}$ because $\sqrt{n}\gamma_{n,jj'}$ is of the same order as $n^{\frac{\xi}{2}}$ while $\sqrt{n}\rho_{jj'}=o(n^{\frac{\xi}{2}})$ by Assumption~\ref{max_rho}.

Consequently, the expected $FPR$ is controlled as desired, 
\begin{align*}
\Mean(FPR) &= \frac{1}{|\mathcal{E}^c|}\sum_{\left(j,j'\right)\notin\mathcal{E}}\Pr\left(\left|\widehat{S}_{jj'}^{\tau}\right|>\gamma_{n,jj'}\right)\\
& \to \frac{\sum_{\left(j,j'\right)\notin\mathcal{E}}\left(\frac{2f}{p(p-1)}\right)}{\left|\mathcal{E}^c\right|}=2f/\left(p\left(p-1\right)\right)\leq f/|\mathcal{E}^c|,
\end{align*}
where the last inequality results from the fact that $|\mathcal{E}^c|\leq\frac{p(p-1)}{2}$.

\bibliographystyle{apalike}
\bibliography{ref}

\begin{thebibliography}{}

\bibitem[Anandkumar et~al., 2012]{an12}
Anandkumar, A., Tan, V. Y.~F., Huang, F., and Willsky, A.~S. (2012).
\newblock {H}igh-dimensional {S}tructure {Estimation} {i}n {I}sing {M}odels:
  {L}ocal {S}eparation {C}riterion.
\newblock {\em The Annals of Statistics}, 40:1346--1375.

\bibitem[Arvesen, 1969]{ar69}
Arvesen, J.~N. (1969).
\newblock {J}ackknifing {U}-{S}tatistics.
\newblock {\em The Annals of Mathematical Statistics}, 40:2076--2100.

\bibitem[Callaert and Veraverbeke, 1981]{cv81}
Callaert, H. and Veraverbeke, N. (1981).
\newblock {T}he {O}rder of the {N}ormal {A}pproximation for a {S}tudentized
  {U}-{S}tatistic.
\newblock {\em The Annals of Statistics}, 9:194--200.

\bibitem[Chen et~al., 2015]{chen15}
Chen, S., Witten, D., and Shojaie, A. (2015).
\newblock {S}election and estimation for mixed graphical models.
\newblock {\em Biometrika}, 102(1):47--64.

\bibitem[Fligner and Rust, 1983]{fr83}
Fligner, M.~A. and Rust, S.~W. (1983).
\newblock {O}n the independence problem and {K}endall's tau.
\newblock {\em Communications in Statistics - Theory and Methods},
  12:1597--1607.

\bibitem[Friedman et~al., 2008]{fr08}
Friedman, J., Hastie, T.~J., and Tibshirani, R.~J. (2008).
\newblock {S}parse inverse covariance estimation with the graphical lasso.
\newblock {\em Biostatistics}, 9:432--441.

\bibitem[Hoeffding, 1963]{hoef63}
Hoeffding, W. (1963).
\newblock {P}robability inequalities for sums of bounded random variables.
\newblock {\em Journal of the American Statistical Association}, 58:13--30.

\bibitem[Lee, 1985]{lee85}
Lee, A.~J. (1985).
\newblock {O}n estimating the variance of a {U}-statistic.
\newblock {\em Communications in Statistics - Theory and Methods}, 14:289--302.

\bibitem[Liu et~al., 2012a]{liu12a}
Liu, H., Han, F., Yuan, M., Lafferty, J., and Wasserman, L. (2012a).
\newblock {H}igh-dimensional semiparametric {G}aussian copula graphical models.
\newblock {\em The Annals of Statistics}, 40:2293--2326.

\bibitem[Liu et~al., 2012b]{liu12b}
Liu, H., Han, F., and Zhang, C. (2012b).
\newblock Transelliptical graphical models.
\newblock In {\em Advances in Neural Information Processing Systems 25}, pages
  809--817.

\bibitem[Liu et~al., 2009]{liu09}
Liu, H., Lafferty, J., and Wasserman, L. (2009).
\newblock {T}he nonparanormal: {S}emiparametric estimation of high dimensional
  undirected graphs.
\newblock {\em Journal of Machine Learning Research}, 10:2295--2328.

\bibitem[Luo et~al., 2015]{luo12}
Luo, S., Shi, C., Song, R., Xie, Y., and Witten, D. (2015).
\newblock {S}ure {S}creening for {G}aussian {G}raphical {M}odels.
\newblock {\em to appear}.

\bibitem[Meinshausen and B{\"u}hlmann, 2006]{mb06}
Meinshausen, N. and B{\"u}hlmann, P. (2006).
\newblock {H}igh-dimensional graphs and variable selection with the lasso.
\newblock {\em The Annals of Statistics}, 34:1436--1462.

\bibitem[Ravikumar et~al., 2009]{ra09}
Ravikumar, P., Lafferty, J., Liu, H., and Wasserman, L. (2009).
\newblock {S}parse additive models.
\newblock {\em Journal of the Royal Statistical Society: Series B (Statistical
  Methodology)}, 71:1009--1030.

\bibitem[Ravikumar et~al., 2010]{ra10}
Ravikumar, P., Wainwright, M.~J., and Lafferty, J.~D. (2010).
\newblock {H}igh-dimensional {I}sing {M}odel {S}election {U}sing
  {L}1-{R}egularized {L}ogistic {R}egression.
\newblock {\em The Annals of Statistics}, 38:1287--1319.

\bibitem[Rothman et~al., 2008]{ro08}
Rothman, A.~J., Bickel, P.~J., Levina, E., and Zhu, J. (2008).
\newblock {S}parse permutation invariant covariance estimation.
\newblock {\em Electronic Journal of Statistics}, 2:494--515.

\bibitem[Sen, 1977]{sen77}
Sen, P.~K. (1977).
\newblock {S}ome {I}nvariance {P}rinciples {R}elating to {J}ackknifing and
  {T}heir {R}ole in {S}equential {A}nalysis.
\newblock {\em The Annals of Statistics}, 5:316--329.

\bibitem[Tan et~al., 2015]{km15}
Tan, K.~M., Witten, D., and Shojaie, A. (2015).
\newblock {T}he cluster graphical lasso for improved estimation of {G}aussian
  graphical models.
\newblock {\em Computational Statistics and Data Analysis}, 85:23--36.

\bibitem[Witten et~al., 2011]{witten2011new}
Witten, D.~M., Friedman, J.~H., and Simon, N. (2011).
\newblock New insights and faster computations for the graphical lasso.
\newblock {\em Journal of Computational and Graphical Statistics},
  20(4):892--900.

\bibitem[Yang et~al., 2012]{NIPS2012_0659}
Yang, E., Allen, G., Liu, Z., and Ravikumar, P. (2012).
\newblock Graphical models via generalized linear models.
\newblock In {\em Advances in Neural Information Processing Systems 25}, pages
  1367--1375.

\bibitem[Yang et~al., 2014]{yang14}
Yang, E., Ravikumar, P., Allen, G., Baker, Y., Wan, Y., and Liu, Z. (2014).
\newblock {A} {G}eneral {F}ramework for {M}ixed {G}raphical {M}odels.
\newblock {\em arXiv:1411.0288}.

\bibitem[Yuan and Lin, 2007]{yl07}
Yuan, M. and Lin, Y. (2007).
\newblock {M}odel selection and estimation in the {G}aussian graphical model.
\newblock {\em Biometrika}, 94:19--35.

\end{thebibliography}
\end{document}